\documentclass[aps,prl,twocolumn,superscriptaddress,preprintnumbers,nofootinbib]{revtex4-1}
\usepackage{amsmath,amssymb,graphicx,natbib,bm}
\usepackage{enumerate}
\usepackage{hyperref}
\usepackage{breakurl}
\usepackage{color}

\newcommand{\fslash}[1]{\!\not\!{#1}}

\def\be{\begin{equation}}
\def\ee{\end{equation}}
\def\ba{\begin{eqnarray}}
\def\ea{\end{eqnarray}}

\usepackage[table]{xcolor}
\usepackage[utf8]{inputenc} % accents in bibtex

\begin{document} 

\title{How long does the hydrogen atom live?}

\author{David McKeen}
\email{mckeen@triumf.ca}
\affiliation{TRIUMF, 4004 Wesbrook Mall, Vancouver, BC V6T 2A3, Canada}

\author{Maxim Pospelov}
\email{pospelov@umn.edu}
\affiliation{School of Physics and Astronomy, University of Minnesota, Minneapolis, MN 55455, USA}
\affiliation{William I. Fine Theoretical Physics Institute, School of Physics and Astronomy, University of Minnesota, Minneapolis, MN 55455, USA}

\begin{abstract}
It is possible that the proton is stable while atomic hydrogen is not. This is the case in models with new particles carrying baryon number which are light enough to be stable themselves but heavy enough so that proton decay is kinematically blocked. Models of new physics that explain the neutron lifetime anomaly generically have this feature, allowing for atomic hydrogen to decay through electron capture on a proton. We calculate the radiative hydrogen decay rate involving the emission of a few hundred keV photon, which makes this process detectable in experiment. In particular, we show that the low energy part of the Borexino spectrum is sensitive to radiative hydrogen decay, and turn this into a limit on the hydrogen lifetime of order $10^{30}~\rm s$ or stronger. For models where the neutron mixes with a dark baryon, $\chi$, this limits the mixing angle to roughly $10^{-11}$, restricting the $n\to\chi\gamma$ branching to $10^{-4}$, over a wide range of parameter space.
\end{abstract}

\date{\today}

\maketitle

%%%%%%%%%%%%%%%%
%%%%%%%%%%%%%%%%

{\it Introduction.}---The stability of matter can be viewed as a consequence of the conservation of baryon number, $B$. In the Standard Model (SM), $B$ is an accidental symmetry: a charge carried by quarks (conventionally normalized to $1/3$), the conservation of which is not imposed by hand but rather results, at the renormalizable level, given the choice of the SM fermion gauge charges. In the effective field theory describing hadrons, there is no good argument why the operator $\pi^0 \bar e^c  p$, with $e^c$ the positron, which would mediate proton decay should be absent. In the context of the SM involving elementary quarks, $q$, and leptons, $l$, their gauge couplings dictate that the lowest dimensional interaction that can give rise to this effective operator is dimension-6, $qqql$, and can therefore be highly suppressed, explaining the apparent stability of the proton (see, e.g., Ref. \cite{Zee:2008bu}). In the effective field theory language, this means that the proton is the lightest particle carrying nonzero $B$ ($B=1$ in the conventional normalization) which is conserved (to at least very good approximation) and thus, the proton is (at least very nearly) stable.

The current constraints on proton decay are somewhat model-dependent, reflecting the experimental capability of detecting the potential final states. The large majority of the considered proton decay modes produce a positron at the end of the decay chain, so that the total energy released in such process, $Q$, is close to the proton mass, although some can be carried away by neutrinos. For the flagship decay mode, $p\to \pi^0e^+$, one can set a very stringent bound, $\tau_p > 5\times 10^{33}~{\rm yr}$~\cite{TheSuper-Kamiokande:2017tit}. The limit benefits from the very distinct signature of this decay in the Super-Kamiokande detector with little background and high efficiency. Somewhat less stringent limits are set by ``disappearing" nucleons in the SNO~\cite{Ahmed:2003sy} and KamLAND experiments~\cite{Araki:2005jt}. These dark decays may include $n\to 3\nu$ modes. The sensitivity to such decay modes comes from detecting the visible energy (e.g. photons) emitted in the process of filling 
the nucleon vacancy in $^{12}$C and $^{16}$O left by the vanished nucleon, so that the {\em visible} $Q$ is on the order of a few MeV. 

In this work, we will consider some aspects of nucleon decay when $Q$ is generally smaller than the nucleon binding energy inside a nucleus. Such decays often result when considering new states with masses close to the nucleon mass scale, which has received considerable attention in recent years. As the simplest example that illustrates the idea underlying our work, consider a toy model involving a dark scalar 
$\phi$ that couples to the proton and electron,
\begin{equation}
    {\cal L} \supset \lambda \bar e^c p \phi +{\rm h.c.},
    \label{eq:toy}
\end{equation}
where $\lambda$ is some coupling constant. Depending on the mass of $\phi$, one can have proton decay, atomic hydrogen decay, or both:
\begin{eqnarray}
m_\phi < m_p -m_e:~~ &&p\to \phi e^+,~~ Q = m_p-m_e-m_\phi, \nonumber\\
m_\phi < m_p + m_e:~~ &&{\rm H} \to \phi  \gamma,~~ Q = m_p+m_e-m_\phi.
\nonumber
\end{eqnarray}
If the scalar mass is in the range $ m_p-m_e < m_\phi < m_p+m_e$, proton decay is not allowed, avoiding the extremely strong bound on the proton lifetime, but the hydrogen atom can decay with the rate
\begin{equation}
\begin{aligned}
    \Gamma_{\rm H\to\phi\gamma}&\simeq\left|\psi(0)\right|^2\frac{\alpha\lambda^2Q}{4m_e^2M}
    \\
    &=\frac{1}{10^{32}~{\rm s}}\left(\frac{\lambda}{10^{-20}}\right)^2\left(\frac{Q}{m_e}\right),
    \end{aligned}
    \label{eq:toyrate}
\end{equation}
where $\psi(0)\simeq\sqrt{\alpha^3 m_e^3/\pi}$ is the value of the electron wavefunction at the location of the proton and $M\simeq m_p+m_e$ is the mass of hydrogen. 
This decay of hydrogen involves the emission of a photon with energy $\omega=Q$ of several 
hundred $\rm keV$ which can be efficiently searched for using the Borexino detector where the
threshold is close to $200~\rm keV$. In particular, the search for the charge-violating 
decay mode, $e\to \gamma + {\rm missing ~energy}$ \cite{Agostini:2015oze}, 
can be directly recast as a constraint on the hydrogen lifetime in model (\ref{eq:toy}) of roughly $\tau_{\rm H} \gtrsim10^{28}~{\rm yr}$. One should note that within this toy model the limits are  
very strong because $\rm H$ decay here {\em must} be accompanied by the emission of a photon. In other models, the leading ${\rm H}\to X$ mode can be fully invisible, with the subdominant radiative ${\rm H}\to X\gamma$ mode involving a photon suppressed by $\alpha$ and an additional phase space factor.

We will also focus on a more motivated scenario than this simple toy model, in which the neutron mixes with a neutral dark fermion, $\chi$. Like the proton's coupling to the electron, since the neutron is a composite state carrying $B=1$, $n$-$\chi$ mixing also occurs at dimension-6 through the operator $qqq\chi$, hence we can justifiably view the mixing as a small parameter. If $\chi$ is a Dirac fermion, one can assign it $B=1$. Curiously, if $m_\chi$ is in the range $m_p-m_e<m_\chi<m_p+m_e$ (note the range for $\phi$ above), $\chi$ and the proton are stable for the same reason: the conservation of $B$~\cite{McKeen:2015cuz}. Since $\chi$ is neutral and stable, it can be considered  as a viable dark matter candidate~\cite{Karananas:2018goc}. The similarity of the $\chi$ and $n$ mass could result from an underlying mirror symmetry~\cite{Berezhiani:2003xm,*Berezhiani:2005hv,*Berezhiani:2005ek,*Berezhiani:2015afa} or be argued for by anthropic reasoning related to the need for dark matter~\cite{McKeen:2015cuz}. Furthermore, if $m_\chi<m_n$, then a new neutron decay mode opens up, $n\to\chi\gamma$, which has been suggested as the solution to the discrepancy between measurements of the neutron lifetime using the ``bottle" and ``beam" methods~\cite{Fornal:2018eol}, also known as the ``neutron lifetime anomaly." However, $n\to\chi\gamma$ decays have been directly searched for and not seen at the level required to explain the discrepancy for $937.8~{\rm MeV}<m_\chi<938.8~{\rm MeV}$~\cite{Tang:2018eln}. We note that this solution to the lifetime anomaly may be in tension with measured neutron $\beta$ decay angular correlations~\cite{Czarnecki:2018okw,*Dubbers:2018kgh}. There are also strong limits from the existence of heavy neutron stars~\cite{McKeen:2018xwc,*Baym:2018ljz,*Motta:2018rxp}. Extensions of the model leave open the possibility of maintaining the dark particle solution to the neutron lifetime anomaly (see, e.g.,~\cite{Cline:2018ami,*Ivanov:2018vit,*Berezhiani:2018eds,*Bringmann:2018sbs,*Grinstein:2018ptl}).

In addition to $n$ decay, hydrogen decay to $\chi$ can occur in this model~\cite{Berezhiani:2018udo} in precisely the mass range where $\chi$ is stable (and thus a potential dark matter candidate), with a radiative branching including a photon of ${\cal O}(\alpha/4\pi)$. In this letter, we will show that data from Borexino can be used to set a stronger limit on the model than the direct search for $n\to\chi\gamma$ over a large range of parameter space where $\rm H$ is destabilized via H$\to \nu\chi$.

%%%%%%%%%%%%%%%%
%%%%%%%%%%%%%%%%
{\it Hydrogen decay.}---We now discuss hydrogen decay in the scenario where the neutron mixes with a new state as well as in a general effective field theory treatment.

The Lagrangian describing the mixing of the neutron with a Dirac fermion $\chi$ carrying $B=1$ is
\begin{equation}
    {\cal L}=\bar n\left(i\fslash{\partial}-m_n\right)  n + \bar \chi\left(i\fslash{\partial}-m_\chi\right)\chi-\delta\left(\bar n \chi + \bar \chi n \right),
    \label{eq:Leff}
\end{equation}
plus terms responsible for weak and electromagnetic interactions of neutrons. 
The mixing strength $\delta$ is empirically required to be small, $\delta\ll m_n,\,m_\chi$. This mass matrix of Eq.~(\ref{eq:Leff}) is diagonalized by taking $n\to n-\theta\chi$, $\chi\to\chi+\theta n$ with the mixing angle given by $\theta=\delta/\Delta m$ with $\Delta m\equiv m_n-m_\chi$.

In this model, to ensure the stability of the proton, $m_\chi>m_p-m_e=937.76~\rm MeV$. Kinematically forbidding the decay $^9{\rm Be}\to\chi\, ^8{\rm Be}$ increases the lower bound on $m_\chi$ by $140~\rm keV$ to $937.900~\rm MeV$~\cite{McKeen:2015cuz} while forbidding the decay to two $\alpha$ particles requires $m_\chi>937.993~\rm MeV$~\cite{Pfutzner:2018ieu}. If $m_\chi<m_p+m_e=938.78~\rm MeV$, the dark baryon $\chi$ is itself stable and thus a potential dark matter candidate~\cite{McKeen:2015cuz,Fornal:2018eol}.

For $m_\chi<m_n$, this model leads to a new decay channel for the neutron, $n\to\chi\gamma$, with branching ratio
\begin{equation}
    {\rm Br}_{n\to\chi\gamma} \simeq 0.02 \left(\frac{\theta}{10^{-9}}\right)^2\left(\frac{\Delta m}{\rm MeV}\right)^3,
\end{equation}
which, as mentioned above, was proposed as an explanation of the neutron lifetime anomaly~\cite{Fornal:2018eol}. For dark baryon masses between $937.8$ and $938.8~\rm MeV$, the direct search for $n\to\chi\gamma$~\cite{Tang:2018eln} limits its branching ratio to ${\cal O}(0.1\%)$.

If $m_\chi<m_p+m_e=938.78~\rm MeV$, as pointed out in Ref.~\cite{Berezhiani:2018udo}, atomic hydrogen can decay through electron capture, $e^-p\to\nu\chi$. The hydrogen decay rate in the presence of $n$-$\chi$ mixing is
\begin{equation}
\begin{aligned}
    \Gamma_{H\to \nu\chi}&=\frac{1}{\tau_H} \simeq\left|\psi(0)\right|^2\frac{G_F^2\left|V_{ud}\right|^2 \theta^2}{2\pi}\left(1+3g_A^2\right)Q^2
    \\
    &=\frac{1}{10^{27}~{\rm s}}\left(\frac{\theta}{10^{-9}}\right)^2\left(\frac{Q}{m_e}\right)^2,
\end{aligned}
\label{eq:GammaHNM}
\end{equation}
where $g_A\simeq 1.27$ is the nuclear axial vector coupling and here $Q=M-m_\chi\ll m_\chi, M$. For $\theta\lesssim 10^{-4}$ and $Q\sim m_e$, the hydrogen lifetime is longer than the age of the Universe and results in a final state that does not interact strongly with normal matter and thus is seemingly difficult to probe.

\begin{figure}
\includegraphics[width=0.8\linewidth]{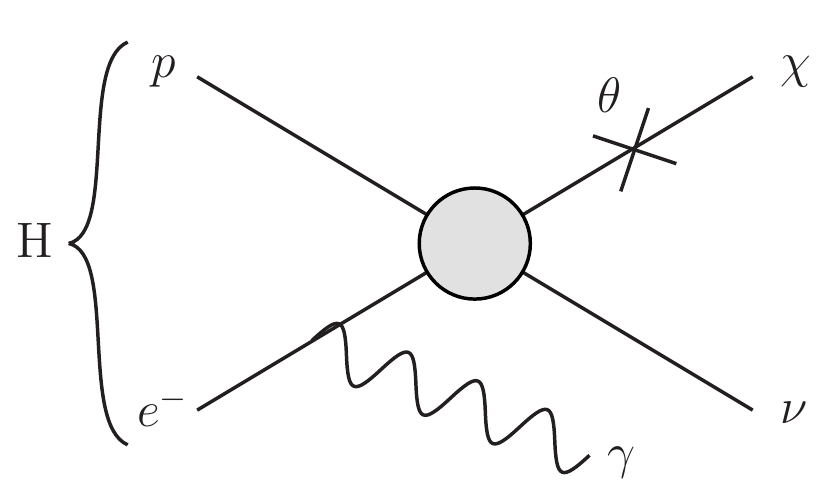}
\caption{Radiative hydrogen decay in the neutron mixing model. A similar diagram can be drawn in the toy model and EFT we consider where we replace $\nu$ and $\chi$ with $\phi$ or $\ell$ and $b$, respectively.}\label{fig:decay}
\end{figure}

In addition to the fully invisible final state, there is a subdominant radiative decay mode, $H\to \nu\chi\gamma$, shown in Fig.~\ref{fig:decay}, which produces a photon with energy $\omega<Q
\sim {\cal O}(100~{\rm keV})$ which can be observed.
The radiative branching fraction in this model as a function of the photon energy is
\begin{equation}
\begin{aligned}
\frac{d}{d\omega}{\rm Br}_{H\to \nu\chi\gamma}&= \frac{\alpha}{\pi}\frac{\omega}{m_e^2}\left(1-\frac{\omega}{Q}\right)^2+{\cal O}\left(\frac{m_e}{m_p}\right)
\\
&\simeq \frac{5\times10^{-6}}{{\rm keV}}\frac{\omega}{m_e}\left(1-\frac{\omega}{Q}\right)^2.
\end{aligned}
\label{eq:BRgammaNM}
\end{equation}
This is peaked at $\omega=Q/3$. The total radiative branching fraction is
\begin{equation}
\begin{aligned}
{\rm Br}_{H\to \nu\chi\gamma}&\simeq \frac{\alpha}{12\pi}\frac{Q^2}{m_e^2}\simeq 2\times10^{-4}\left(\frac{Q}{m_e}\right)^2.
\end{aligned}
\end{equation}

Other models with sub-GeV states carrying $B$ and lepton number, $L$, can lead to the decay of hydrogen. Such models have often been considered in the context of asymmetric dark matter (see, e.g.~\cite{Kaplan:1991ah,*Kitano:2004sv,*Kaplan:2009ag,*An:2009vq,*Shelton:2010ta,*Davoudiasl:2010am,*Falkowski:2011xh}). As mentioned above, in the toy model of Eq.~(\ref{eq:toy}), if $m_\phi<M$ then $\rm H\to\phi\gamma$ proceeds emitting a monochromatic photon of energy $\omega=Q=M-m_\phi$.

We also consider a simple setup involving two exotic, light neutral fermions,  $\ell$ and  $b$, carrying $L=1$ and $B=1$, respectively. These can interact with the electron and proton through a dimension-6 operator. For simplicity, we consider the scalar-scalar operator,
\begin{equation}
    {\cal L}_{\ell b}=\frac{1}{\Lambda^2}  (\bar b\, p) (\bar \ell\, e).
    \label{eq:LEFT}
\end{equation}
Of course, depending on the UV completion, other Lorentz structures are possible. Also depending on the UV completions are the strengths of the operators that could induce related processes such as neutron decay. For instance, given the scalar in Eq.~(\ref{eq:toy}), one might expect $\bar\nu^c n\phi$ to exist, while $(\bar b\, n) (\bar \ell\, \nu)$ could exist in the EFT. These operators are necessarily generated in the presence of~(\ref{eq:toy}) and~(\ref{eq:LEFT}) at loop level, but because of the model dependence of their strengths, we do not discuss them further.

As in the neutron-mixing model, the stability of $^9$Be (and the proton) is ensured if $m_\ell+m_b>937.993~\rm MeV$. If $m_\ell+m_b=M-Q<M$, $H\to\ell b$ occurs through~(\ref{eq:LEFT}). We assume that $\phi$, $\ell$, and $b$ are either stable on the scale of the Borexino experiment or that they decay into dark sector states so that they do not leave any other visible signature. 
If $m_b$ or $m_\ell$ is zero, then the rate for this decay is parametrically the same as in Eq.~(\ref{eq:GammaHNM}). When $m_\ell$ and $m_b$ are comparable, the rate scales differently with $Q$. Taking $m_\ell=m_b$ for definiteness, the decay rate is 
\begin{equation}
\begin{aligned}
    \Gamma_{H\to\ell b}&\simeq \left|\psi(0)\right|^2\frac{\sqrt{M^3Q}}{4\sqrt2\pi\Lambda^4}
    \\
    &=\left(3\times 10^{27}~{\rm s}\right)^{-1}\sqrt{\frac{Q}{m_e}}\left(\frac{100~\rm PeV}{\Lambda}\right)^4.
\end{aligned}
\label{eq:tHequal}
\end{equation}

For $m_\ell=0$ the radiative branching ratio is the same as in the neutron-mixing case in~(\ref{eq:BRgammaNM}) while for $m_b=0$ it is simply half that. Given the similarity of the $m_{\ell,b}=0$ rates to the neutron-mixing case, we will not consider these points in parameter space further, noting that limits can simply be translated from the neutron-mixing case. 

For equal masses, $m_\ell=m_b$, the photon spectrum in radiative $\rm H$ decay is slightly harder than in neutron mixing, peaked at $\omega=2Q/3$,
\begin{equation}
\begin{aligned}
    \frac{d{\rm Br}_{H\to\ell b\gamma}}{d\omega}&\simeq\frac{\alpha}{2\pi}\frac{\omega}{m_e^2}\sqrt{1-\frac{\omega}{Q}}.
\end{aligned}
\end{equation}
We now move on to discuss the experimental signature of these scenarios from radiative hydrogen decay.

{\it Decays at Borexino.}---The Borexino solar neutrino experiment contains a large amount of radio-pure organic scintillator, and thus hydrogen atoms, in a low-background environment.  Its extreme radio-purity allows the threshold for the detection of electromagnetic energy depositions to be reduced down to $\sim 200~\rm keV$ set by the $^{14}$C background. It is therefore the most promising experiment to search for the radiative decay of hydrogen. The fiducial volume of Borexino is ${\cal O}(100~{\rm t})$ of pseudocumene which is about 10\% hydrogen by weight. This means that, in the neutron-mixing model, the total radiative hydrogen decay rate at Borexino is about
\begin{equation}
    \frac{4\times 10^{4}}{100~{\rm t}\,{\rm day}}\left(\frac{\theta}{10^{-9}}\right)^2\left(\frac{Q}{m_e}\right)^4\left(\frac{f_{\rm mol}}{0.5}\right),
\end{equation}
where the last factor,
\begin{equation}
    f_{\rm mol}\equiv\left|\frac{\psi_{\rm mol}(0)}{\psi(0)}\right|^2,
\end{equation}
represents the reduction in the probability of finding the electron at the location of the proton in the molecular state from that in atomic hydrogen. We somewhat conservatively normalize this to $0.5$; note that the value of $f_{\rm mol}$ in simple hydrocarbons, e.g., methane~\cite{doi:10.1063/1.1730858,*doi:10.1063/1.1734482,*doi:10.1246/bcsj.46.1428}, can be slightly larger than this.

In the scalar toy model, since the radiative mode is the leading decay mode and the photon is emitted monochromatically, it is more physically meaningful to parameterize the number of radiative decays simply by the $\rm H$ lifetime than by $\lambda$. The number of events expected at Borexino is then
\begin{equation}
    \frac{3\times 10^{3}}{100~{\rm t}\,{\rm day}}\left(\frac{10^{32}~\rm s}{\tau_{\rm H}}\right).
\end{equation}

For the effective operator of~(\ref{eq:LEFT}) with  $m_\ell=m_b$, the total photon production rate is roughly
\begin{equation}
    \frac{2\times 10^{4}}{100~{\rm t}\,{\rm day}}\left(\frac{100~\rm PeV}{\Lambda}\right)^4 \left(\frac{Q}{m_e}\right)^4\left(\frac{f_{\rm mol}}{0.5}\right).
\end{equation}
Given that the total rate of electromagnetic energy deposition seen at Borexino above $225~\rm keV$ is about $600/(100~{\rm t~day})$, these rough estimates make it clear observable event rates are possible for $\theta\sim10^{-9}$ in neutron-mixing, $\tau_H\sim10^{32}~\rm s$ in the toy model of (\ref{eq:toy}), and $\Lambda\sim 100~\rm PeV$ in (\ref{eq:LEFT}) as long as $Q$ is larger than $\sim 225~\rm keV$.

\begin{figure}
\includegraphics[width=\linewidth]{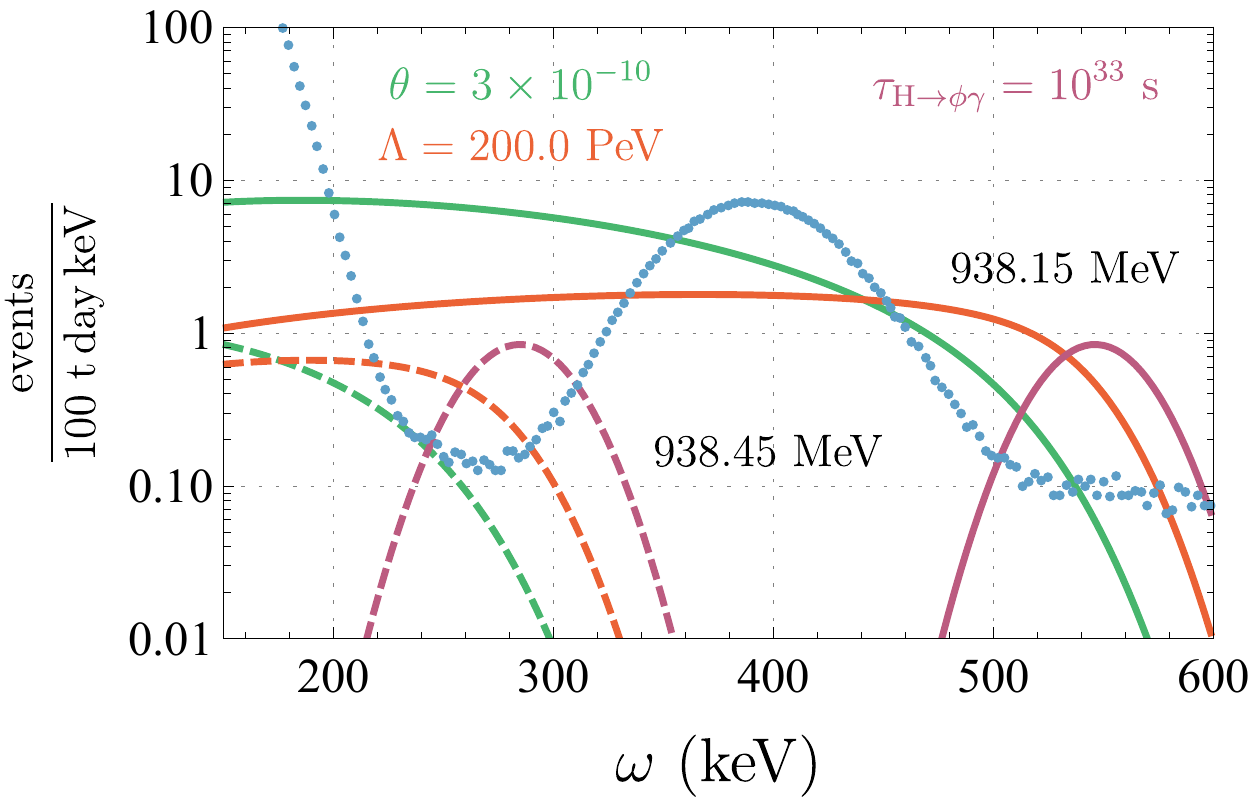}
\caption{Data from Borexino~\cite{Agostini:2015oze} (blue points) along with the photon spectra from radiative hydrogen decay in neutron-mixing (green) with $\theta=3\times10^{-10}$, in the EFT with $m_\ell=m_b$ (red) and $\Lambda=200~\rm PeV$, and in the toy model (purple) with $\tau_{\rm H}=10^{33}~\rm s$. We have chosen $M-Q=m_\chi,m_\ell+m_b,m_\phi=938.15~\rm MeV$ (solid curves) and $938.45~\rm MeV$ (dashed curves). See text for details about the modelling of the detector response.}\label{fig:spectrum}
\end{figure}

In Fig.~\ref{fig:spectrum}, we show the photon spectra at Borexino in the neutron-mixing model, the EFT with $m_\ell=m_b$ and the toy model of Eq.~(\ref{eq:toy}) for $M-Q=938.15~\rm MeV$ and $938.45~\rm MeV$ along with the Borexino data of Ref.~\cite{Agostini:2015oze}. We fix $\theta=3\times 10^{-10}$, $\Lambda=200~\rm PeV$, and $\tau_{\rm H}=10^{33}~\rm s$ in the three models, respectively. To mock up the detector response to photons described in~\cite{Agostini:2015oze}, we assume that the photon energies are quenched by a factor of $0.86$ and smeared according to a gaussian with a full width at half maximum of $50~\rm keV$, and we take a detection efficiency of 25\%. We set 10\% by weight of the detector to be composed of hydrogen and take the square of the molecular electron wavefunction at the proton to be reduced from the atomic value by the factor $f_{\rm mol}=0.5$.

Given the agreement of the Borexino data with expectations of solar neutrinos and backgrounds from radioactivity, we can use these data to determine limits on the allowed values of $\theta$ in neutron-mixing, $\lambda$ (equivalently just $\tau_{\rm H}$) in the toy model of Eq.~(\ref{eq:toy}), or $\Lambda$ in the EFT of Eq.~(\ref{eq:LEFT}). To do so, we fit the measured spectrum with background components from solar neutrinos, decays of $^{14}{\rm C}$, $^{210}{\rm Bi}$, and $^{210}{\rm Po}$, as well as from pileup. As in~\cite{Agostini:2015oze}, we require that the solar neutrino and $^{14}{\rm C}$ rates agree with independent determinations within errors and allow the $^{210}{\rm Bi}$ and $^{210}{\rm Po}$ contributions to float. We then add in signal for a fixed $m_\chi$, $m_\ell+m_b$, $m_\phi=M-Q$ (with efficiency, energy quenching and smearing, and molecular electron wavefunction as described above) and determine the value of $\theta$, $\Lambda$, or $\lambda$ at which $\Delta\chi^2=2.71$, corresponding to a 90\% C.\,L. upper limit. To validate our procedure, we verify that we obtain a similar limit on the injection of monochromatic $256~\rm keV$ photons from $e^-$ decay as in~\cite{Agostini:2015oze}. 

\begin{figure}
\includegraphics[width=\linewidth]{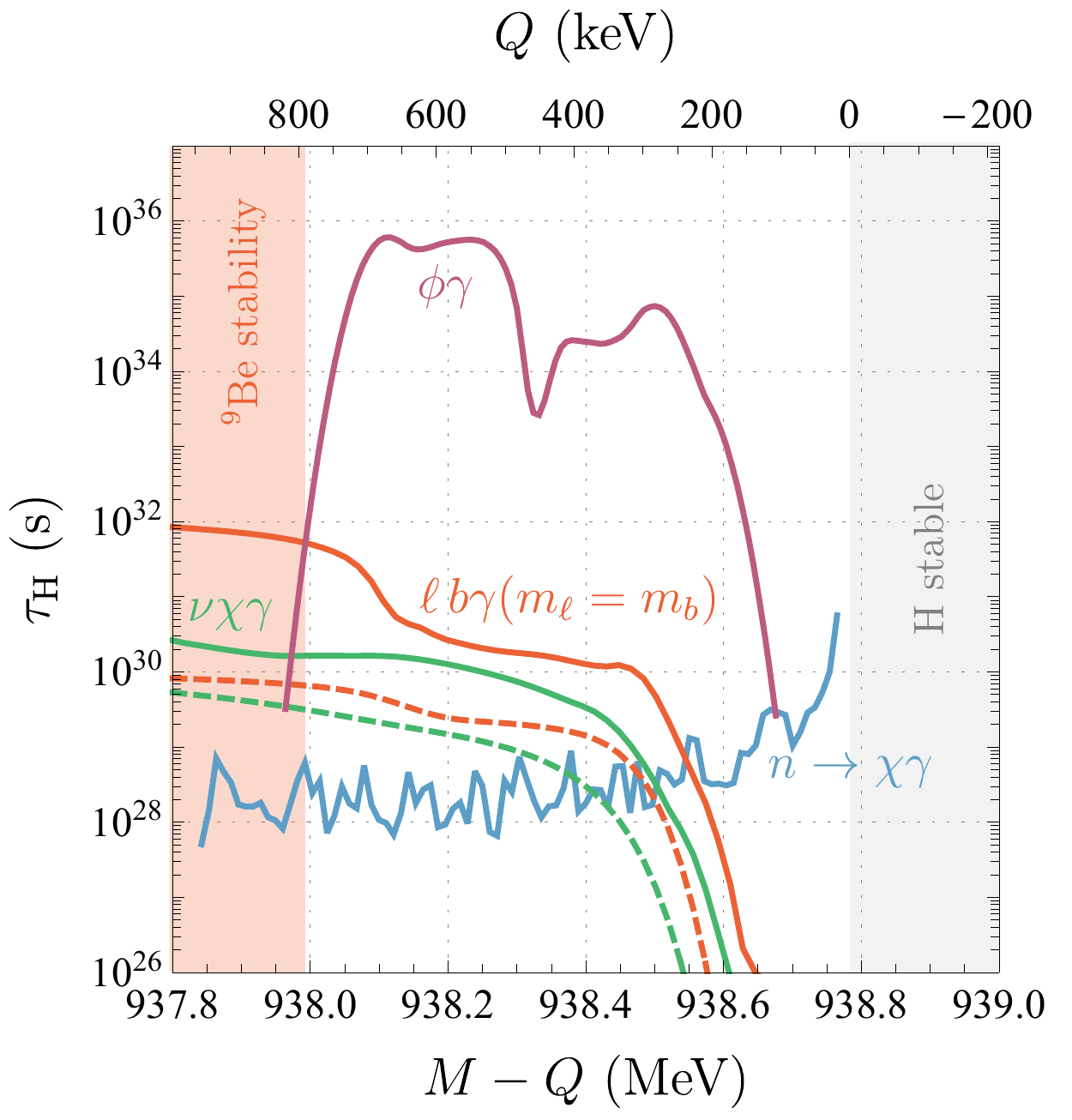}
\caption{Our 90\%~C.\,L. lower limits on the free hydrogen lifetime labeled by the final state in radiative $\rm H$ decay. We show the limit in the EFT with $m_\ell=m_b$ (solid red), the neutron-mixing case (solid green), and the toy model of Eq.~(\ref{eq:toy}) (solid purple) as functions of $M-Q=m_\ell+m_b,m_\chi,m_\phi$. The dashed curves show conservative limits derived from simply requiring the number of signal events with $225~{\rm keV}\le \omega \le 300~{\rm keV}$, $\omega\ge 500~{\rm keV}$ not exceed the observed number. We also show the inferred limit on the neutron-mixing model from the lack of observation of $n\to\chi\gamma$ decays in Ref.~\cite{Tang:2018eln} (solid blue) . The red shaded region is where $^9{\rm Be}$ is unstable and in the gray shaded region hydrogen is stable.}\label{fig:tH}
\end{figure}

We express the upper limits on the number of events as lower limits on the $\rm H$ lifetime in each of these three scenarios which we show in Fig.~(\ref{fig:tH}). The solid lines show the 90\%~CL lower limits from the fit procedure described above for the neutron-mixing (green), toy (purple), and EFT (red) scenarios. The dashed lines are conservative limits on the lifetime that come from requiring that the number of signal events with $225~{\rm keV}<\omega<300~{\rm keV}$ or $\omega>500~\rm keV$ not exceed the total number measured in this range for neutron-mixing (green) and the EFT (red). We also show our estimate of the 90\%~CL lower limit that applies to the neutron-mixing scenario from the direct search for $n\to\chi\gamma$ in Ref.~\cite{Tang:2018eln}, assuming no other exotic decay modes of the neutron. Note that the weakening of the limit on the toy model for $m_\phi\lesssim 938.1~\rm MeV$ comes from our not using data with $\omega>600~\rm keV$.

\begin{figure}
\includegraphics[width=\linewidth]{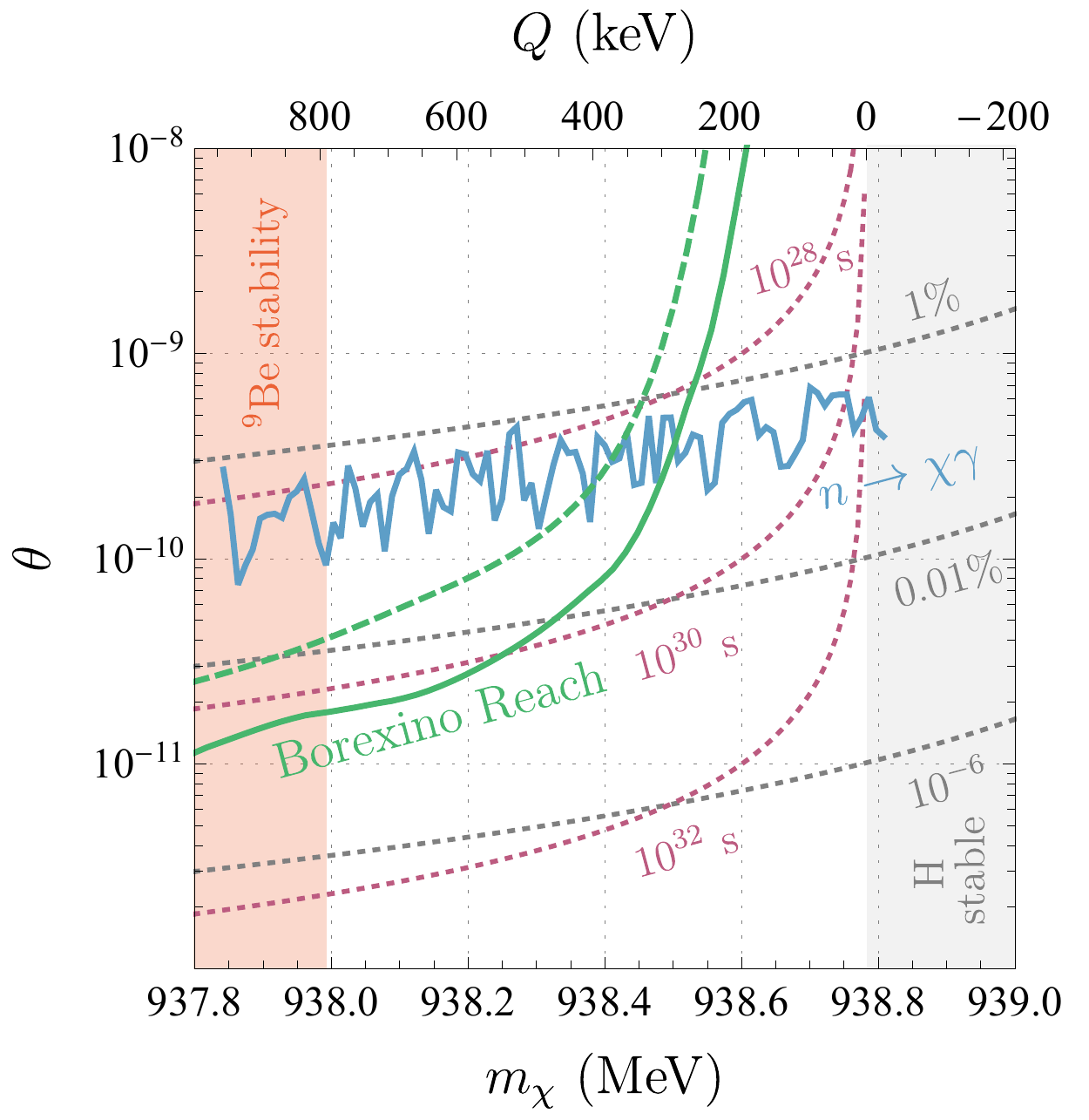}
\caption{The 90\% CL upper limit on $\theta$ as a function of $M-Q=m_\chi$ in the neutron-mixing model from Borexino data~\cite{Agostini:2015oze} (solid green) along with a simple conservative limit as described in Fig.~\ref{fig:tH} (dashed green). Our estimate of the 90\% CL upper limit on $\theta$ from the search for ${\rm Br}_{n\to\chi\gamma}$ (assuming no other exotic neutron decay mode) from~\cite{Tang:2018eln} is shown for comparison (solid blue). The dashed gray contours show ${\rm Br}_{n\to\chi\gamma}$ branching ratios of 1\%, 0.01\%, and $10^{-6}$ and the dashed purple contours indicate atomic hydrogen lifetimes of $10^{28}$, $10^{30}$, and $10^{32}~\rm s$. The red region below $937.993~\rm MeV$ is ruled out by the stability of $^9{\rm Be}$ and in the gray area above $m_p+m_e$, hydrogen does not decay.}\label{fig:limit}
\end{figure}

In addition, we show our upper limit on $\theta$ in the neutron-mixing model in Fig.~\ref{fig:limit}. The solid and dashed curves are computed as in Fig.~\ref{fig:tH} and our estimate of the 90\%~C.\,L. upper limit from Ref.~\cite{Tang:2018eln}.

As can be seen in Figs.~\ref{fig:tH} and~\ref{fig:limit}, the Borexino data can probe $\theta\gtrsim10^{-10}$ and $\Lambda$ as large as $\sim100-1000~\rm PeV$ in the EFT. The limits are stronger in the EFT than for neutron-mixing since the photon spectrum is slightly harder (for $m_b=m_\ell$) which moves the signal out from under the $^{14}{\rm C}$ background into a region with fewer events.

We briefly mention other possible probes of hydrogen stability
for $Q<200$\,keV. 
In the range of $\sim 30-200$ keV, meaningful limits 
can be set from the studies of the $^{14}$C signal from the precursor of the Borexino experiment \cite{Alimonti:1998rc}. Conservatively requiring radiative $\rm H$ decays not exceed the $^{14}$C measured decay rate we arrive at 
\begin{equation}
    {\rm Br}_{{\rm H}\to X\gamma}\times \tau_{\rm H}^{-1} < 10^{-29}\, {\rm s}^{-1}.
\end{equation}
In principle, further studies of $^{14}$C-poor hydrocarbon radioactivity could set limits on H decays in the remaining window for $Q$ down to a few keV. Alternative probes are provided by cosmology, where the injection of energy due to radiative H decays would alter the pattern of cosmic microwave background angular anisotropies. Adopting the results of Ref. \cite{Slatyer:2016qyl}, one can arrive at sensitivity to 
$(Q/m_p){\rm Br}_{{\rm H}\to X\gamma}\tau_{\rm H}^{-1}$ at the level of $10^{-24}\,{\rm s}^{-1}$, which is not competitive in strength with limits derived in this paper.

{\it Conclusions.}---There has been a large amount of recent interest in models involving new states with masses close to the proton and neutron; strong motivations come from the neutron lifetime anomaly and dark matter. While the proton (and $^9{\rm Be}$) lifetime has long been appreciated as a constraint on models of new physics, the fact that neutral hydrogen can decay in such scenarios has not received as much attention. The typical final state for hydrogen decay in such a model is fully invisible which makes it difficult to test. We have shown in this letter that nontrivial constraints can arise from the subdominant radiative decay mode where a photon is also emitted when comparing against data collected at Borexino. This could provide a direct test of scenarios where the neutron mixes with a dark state without the nuclear physics complications that have to be confronted when searching for it in the decay of heavier nuclei such as $^{11}{\rm Be}$~\cite{Pfutzner:2018ieu,Ejiri:2018dun,*Keung:2019wpw}.

In addition, we have also shown limits on more general models, the toy model in Eq.~(\ref{eq:toy}) and dimension-6 EFT in Eq.~(\ref{eq:LEFT}), that can give rise to the decay of hydrogen. In the former, the constraints are quite strong since the leading decay mode involves a photon, while in the latter the limits are comparable to those in the neutron-mixing model since the radiative mode occurs with branching fraction ${\cal O}(\alpha/4\pi)$. In all cases, the limits on the $\rm H$ lifetime attained here are far stronger than those that come from cosmological observations. It would be worthwhile to estimate the limits on $\tau_{\rm H}$ that could be extracted at future experiments, possibly with other organic scintillators or by doping sensitive detectors with hydrogen-containing compounds.

It is amusing that the stability of the proton and atomic hydrogen can be decoupled from one another. The Borexino bounds that we have derived set important constraints for models that are motivated to explain the neutron lifetime anomaly and explain the existence of dark matter. Furthermore, beyond these particular applications, it is interesting on general grounds to quantitatively address the stability of neutral hydrogen itself, which is, after all, the dominant form of atomic matter in our universe.

{\it Acknowledgements.}---We thank Pietro Giampa, David Morrissey, and Nirmal Raj for helpful discussions and Christopher Morris for guidance on Ref.~\cite{Tang:2018eln}. D.\,M. is supported by a Discovery Grant from the Natural Sciences and Engineering Research Council of Canada (NSERC). TRIUMF receives funding through a contribution agreement with the National Research Council of Canada (NRC).

\bibliography{hdecay}

\end{document}